# A NOVEL APPROACH FOR PASS WORD AUTHENTICATION USING BRAIN -STATE -IN -A BOX (BSB) MODEL

A S N Chakravarthy[#1] Penmetsa V Krishna Raja [#2], Prof. P S Avadhani [#3]

*[#1]Department of CSE& IT, Sri Aditya Engineering College, Surampalem, Andhra Pradesh, India,*

*[#2] JNTUK, Kakinada, Andhra Pradesh, India,*

*[#3]Dept. of CS & SE, Andhra University, Visakhapatnam, Andhra Pradesh, India*

***Abstract*— Authentication is the act of confirming the truth of an attribute of a datum or entity. This might involve confirming the identity of a person, tracing the origins of an artefact, ensuring that a product is what it's packaging and labelling claims to be, or assuring that a computer program is a trusted one. The authentication of information can pose special problems (especially man-in-the-middle attacks), and is often wrapped up with authenticating identity. Password authentication using Brain-State -In-A Box is presented in this paper .Here in this paper we discuss Brain-State -In-A Box Scheme for Textual and graphical passwords which will be converted in to probabilistic values Password. We observe how to get password authentication Probabilistic values for Text and Graphical image. This study proposes the use of a Brain-State -In-A Box technique for password authentication. In comparison to existing layered neural network techniques, the proposed method provides better accuracy and quicker response time to registration and password changes.**

## INTRODUCTION

This paper introduces new technique to overcome the limitations of present password techniques. So before introducing this approach of authentication let us introduce some basics of Brain-State-in -Box (BSB).

### *A. Brain -State -In -A Box Model*

The "Brain-State-In-A-Box" [1, 2] (BSB) model is one of the earliest Dynamic Associative Memories (DAM) models. It is a discrete-time continuous-state parallel updated DAM. The BSB model extends the Linear Associator model and is similar to the Hopfield Model in that it is an Auto-associative model with its connection matrix computed using outer products in the usual way. The operation of both models is also very similar, with differences arising primarily in the way activations are computed in each iteration, and in the signal function used. The BSB model stands apart from other models in its use of the linear threshold signal function.

- **Activation Pattern:** $x(t) = [x_1(t), \ldots, x_d(t)]$
- **BSB Algorithm (W,b,$\gamma$):** $x(t) \rightarrow x(t+1)$ where

$$x_i(t+1) = \begin{cases} 1 & \text{if } u_i(t) > +1 \\ u_i(t) & \text{if } |u_i(t)| \le 1 \\ -1 & \text{if } u_i(t) < -1 \end{cases} \quad \text{and} \quad u_i(t) = x_i(t) + \gamma\left(\sum_{j=1}^{d} w_{ij} x_j(t) + b_i\right)$$

- **Connection Matrix:**

$$\mathbf{W} = \begin{bmatrix} w_{11} & \cdots & w_{1d} \\ \vdots & w_{ij} & \vdots \\ w_{1d} & \cdots & w_{dd} \end{bmatrix}$$

$$w_{ij} = w_{ji} \ (\text{symmetric } \mathbf{W})$$

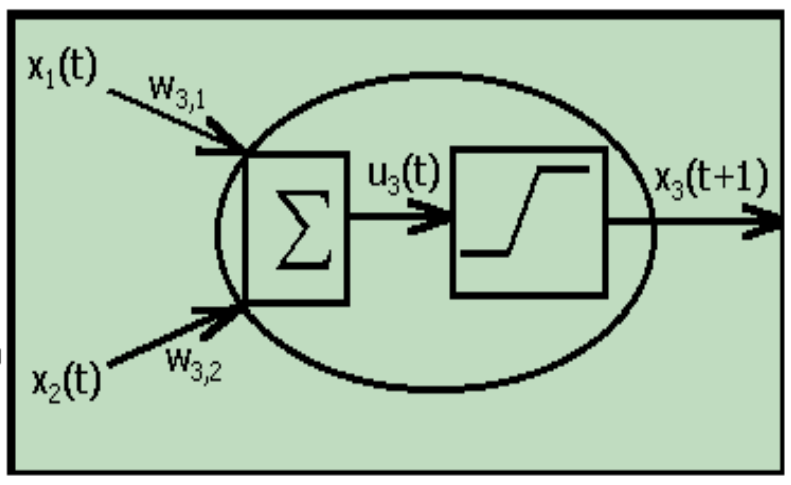

Fig. 1 Working of BSB

## I. USER AUTHENTICATION USING BSB

The architecture of the brain state in a box (BSB) consists of one layer of units that connect to themselves as illustrated in Figure1.The connection weights between units are bidirectional and symmetric. The units may be fully connected, as illustrated in the figure, or only partially connected by randomly setting some of the weights to 0. Anderson and his colleagues have frequently used 50%, or less, connectivity. Partial connectivity does not qualitatively affect the network performance, but reduces computational time, and provides some increase in biological realism (Anderson, 1995).

### *A. Process of Authentication using BSB*

The process of authentication uses any one among Textual password or Graphical password as password and the training will be given to the network so that it can authenticate users.

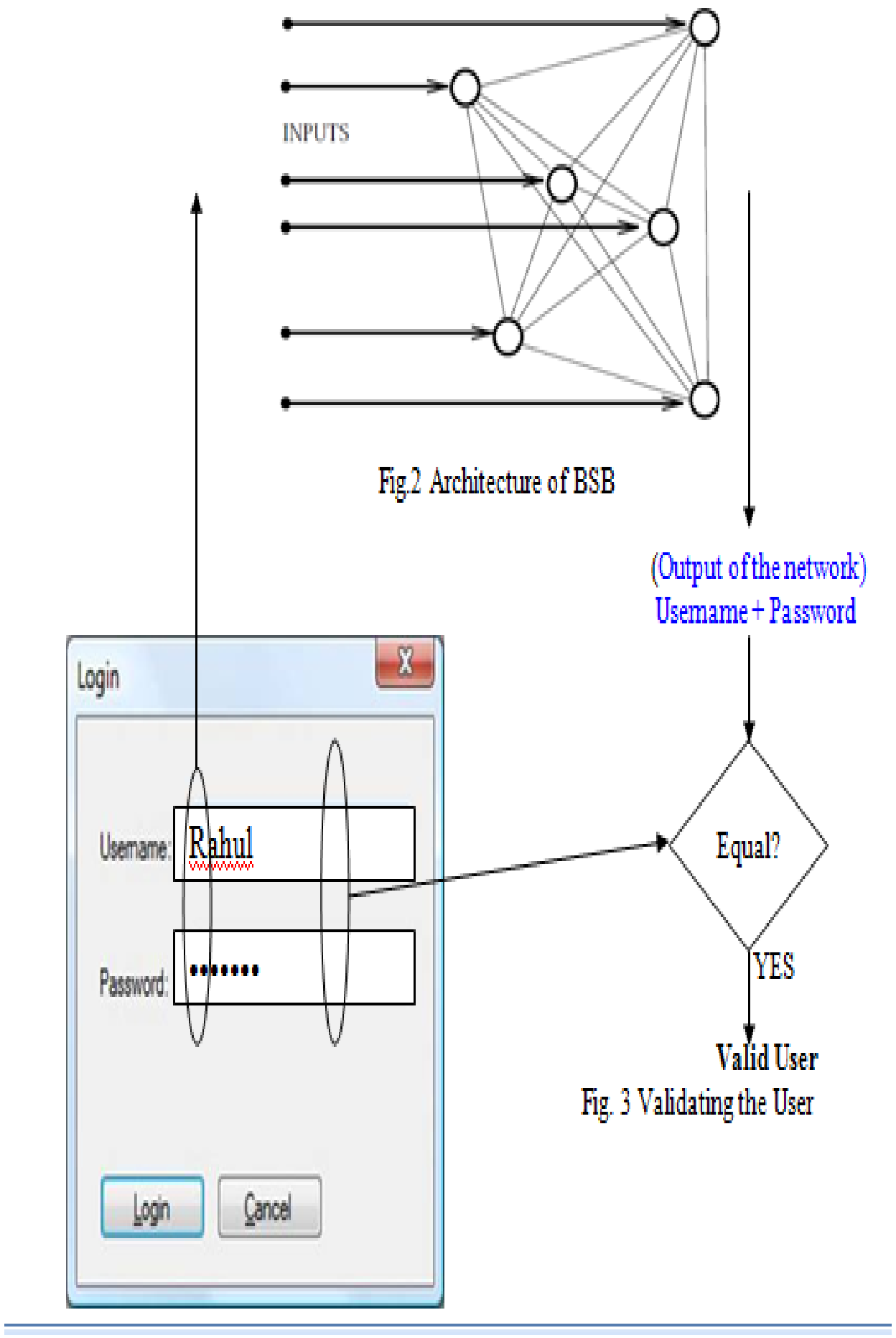

Fig.2 Architecture of BSB

Fig. 3 Validating the User

### B. *Text password*

First this method converts the username and password into binary values and the uses those values as training samples, which can be performed by the following steps

- Convert each character into a unique number (for example ASCII value)
- Convert the unique number into binary value

C

↓

67 (ASCII value)

↓

01000011 (Binary Equivalent of 65)

Fig. Converting Character in to Binary Values

By using above procedure we can convert all characters in the username and password into binary values.

TABLE 1.
BINARY VALUES FOR THE GIVEN USER NAME.

| Username | Binary value representing username |
|---|---|
| SASTRY | 011001010100000101100101000101010010010101001101 |
| VAMSY | 0011010101000001010110010110010101001101 |
| SAI | 011001010100000101001001 |

After converting username and password into binary equivalents the pairs can be used as training samples. Once the training has been completed very soon the network will be stored in each server. When the user wants to get service from a server he/she submits user name and password to the server, then server loads network and generates output by giving username as input. If the output matches with the password submitted by the user then server provides service.

The method can be enhanced for better authentication by using bipolar input instead of binary input. We can convert a binary number into bipolar number by using following formula or by simply replacing zeros with -1s.

If X is a binary digit then corresponding bipolar value is (2X-1).

1 → 1
0 → -1

The above procedure will reinforce in converting binary value in to bipolar value and can be used it as input to the network.

TABLE 2.

BIPOLAR VALUES FOR THE GIVEN USER NAME

| Username | Binary value representing username |
|---|---|
| SASTRY | -111-1-11-11-11-1-1-1-1-11-111-1-11-11-1-1-11-11-11-1-11-1-11-11-11-1-111-11 |
| VAMSY | -1-111-11-11-11-1-1-1-1-11-11-111-1-11-111-1-11-11-11-1-111-11 |
| SAI | -111-1-11-11-11-1-1-1-1-11-11-1-11-1-11 |

### C. *BSB Learning*

Whenever new users are creating accounts network has to adjust weights so that it can recognize all the users who are registered. This process of changing weights is called learning.

### D. *Learning in Brain State in a Box Model*

First, the learning phase establishes the weights for each connection between units of the auto-associative memory. Using either standard Hebbian or Windrow-Hoff techniques, each representation of a multiplication number fact is associated with itself.

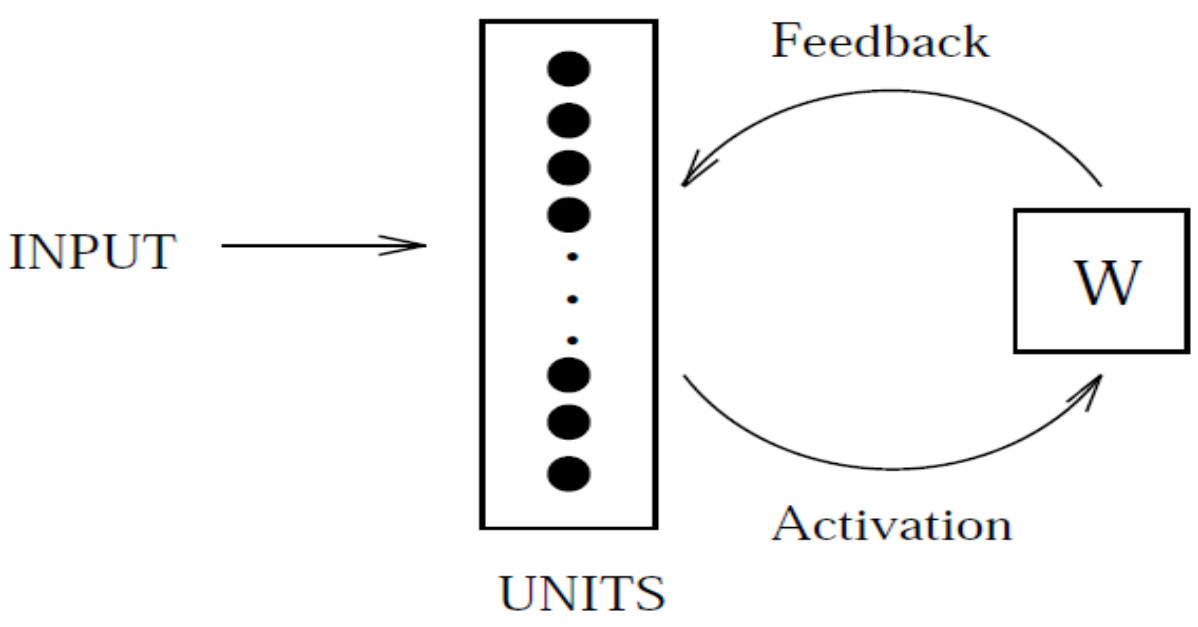

Fig. 5 Implementation details

```
int NoOfPatterns = 0;
int NoOfBitsPerPattern = 0;
int[,] Weight,Input;
```

In this method NoOfPatterns specifies the number of patterns to use in training, ***NoOfBitsPerPattern*** specifies number of bits to use for each pattern, *Weight* stores weight values of the network, Input Stores input vector and *Implementing Training*. Before training, application takes the training samples from the user and stores them in the corresponding variables.

In Training, the NxN connection matrix A is modified as

$$\Delta A = lr * (X - AX) \otimes X \quad (1)$$

$$A = A + \Delta A \quad (2)$$

Where

- X is the normalized input training pattern;
- lr is the Learning rate;
- $\otimes$ is the outer product of two vectors;

```
private void Train()
{
   for (int i = 0; i < Input.GetLength(0); i++)
   {
      int[,] pattern = MatrixMath.GetRow(Input, i);
            int[,] temp1 = MatrixMath.Multiply(Weight,
pattern);
            int[,] temp2 = MatrixMath.Subtract(pattern,
temp1);
            int[,] dw = MatrixMath.Multiply(temp2,pattern);
```

```
        dw = MatrixMath.ScalarMultiply(lr,Weight);
        Weight=MatrixMath.Add(Weight,
ContributionMatrix);
    }
}
```

*E. Recognizing the Pattern using BSB*

Here the pattern which is used for testing the network will be supplied as input to the application and then application stores the pattern in the corresponding variable. The equation 3 is used to calculate the output of the network.

$$X_{[n+1]} = f(\gamma X_n + \eta W X_{[n]} + \delta X) \quad (3)$$

Wher f(x) is defined as follows.

$$f(x) = \begin{cases} -1, & for \quad x < -1 \\ x, & for \quad -1 \le x \le 1 \\ +1, & for \quad x > +1 \end{cases} \quad (4)$$

**This is implemented as follows**

```
private void Recognize()
{
  try
  {
    int[,] pattern = CreateMatrix(tableLayoutPanel2);
    int[,] temp1 =
MatrixMath.ScalarMultiply(gama,pattern);
                int[,] temp2 =
MatrixMath.ScalarMultiply(lr,Weight);
    int[,] temp3 =
MatrixMath.ScalarMultiply(delta,pattern);
    result = MatrixMath.Add(temp1+temp2+temp3);
                result = f(result);

MessageBox.Show(MatrixMath.GetString(result));
  }
  catch (Exception ee)
{ MessageBox.Show(ee.Message);
  }
}
```

## II. BSB for Graphical Password

Image to Bipolar Conversion

By using above procedure first convert the image into matrix representing binary values. Now convert the binary values into bipolar values by replacing 0 with -1 and represent in the form of a matrix.

1 0 0 0 0 1 1 1 1 1 0 0 1 1 1 0 1 1 1 0 1 0 1 1

↓

1 -1 -1 -1 -1 1 1 1 1 1 -1 -1 1 1 1 -1 1 1 1 -1 1 -1 1 1

The above procedure converts the matrix consisting of binary values into a matrix consisting of bipolar values representing all the pixels of the image.

1 -1 -1 -1 -1 1 1 1 1 1 -1 -1 1 1 1 -1 1 1 1 -1 1 -1 1 1 ..............................

.....................................................................................

.....................................................................................

.....................................................................................

1 -1 -1 1 -1 1 -1 -1 ...........................................................

Fig. 6 Conversion of Image to Bipolar

After converting the image into bipolar values the same procedure which is used for textual password can be applied.

## III. Results

*A. BSB for Textual Passwords*

Here *No of Patterns*, specifies the number of patterns application used in the BSB training, *No of Bits Per Pattern* specifies number of bits to be used for each pattern

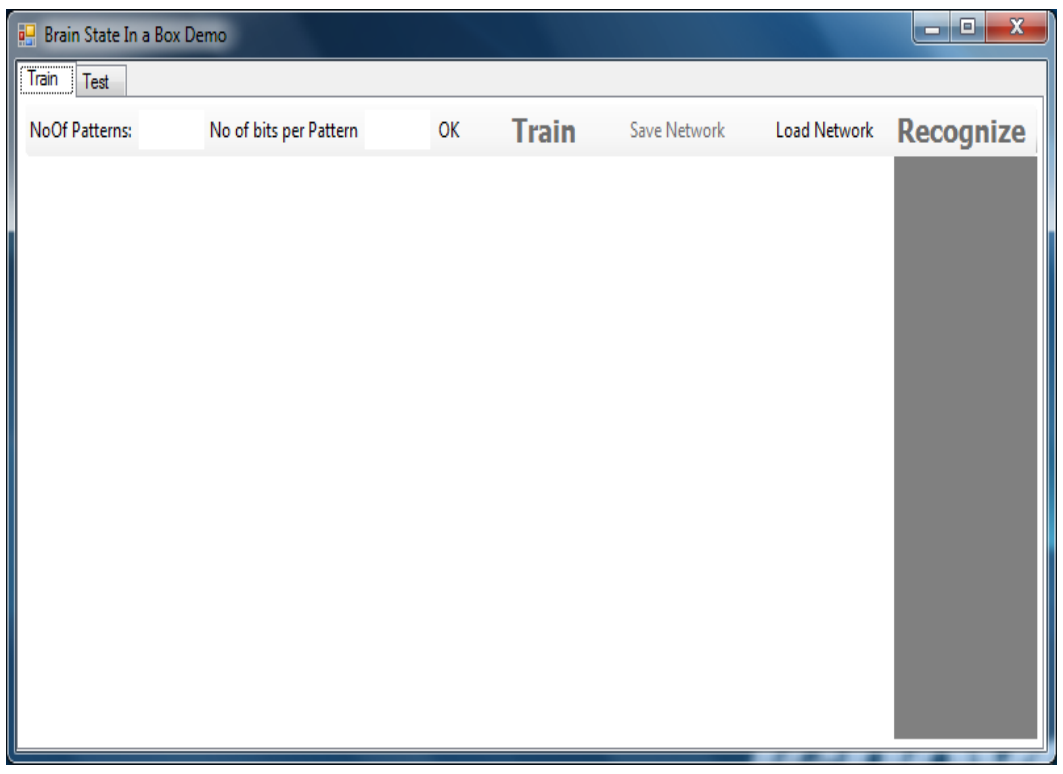

Fig. 7 Screen showing how to setup network

Once the required information has given and OK button is pressed, the application will provide enough fields to enter input and output pairs.

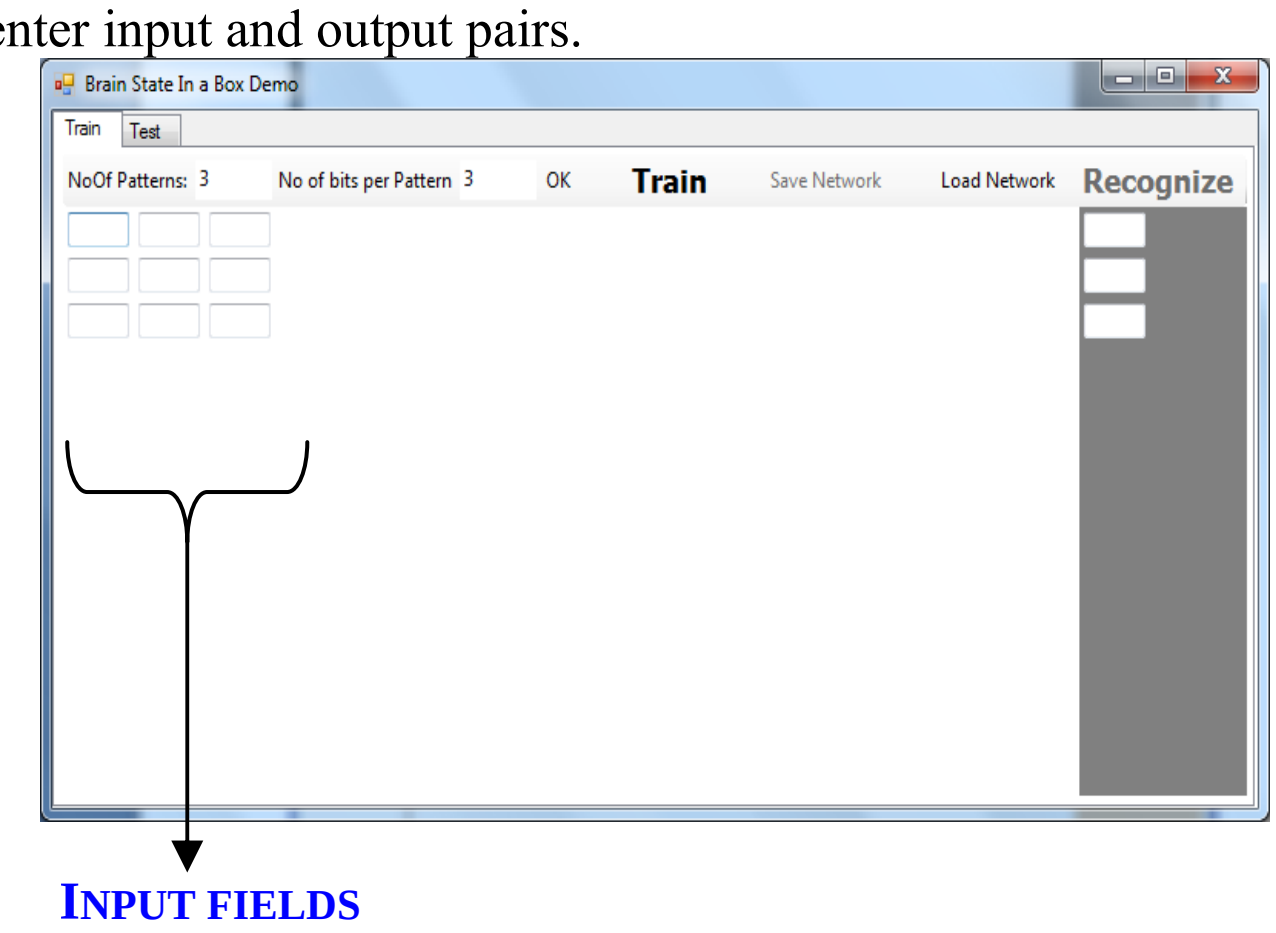

Fig. 8 Screen showing Introduction of BSB

## A. *Training in BSB*

The required training set has given for the application as shown in the figure 9 .When the *Train* button *is* pressed the application will perform training internally.

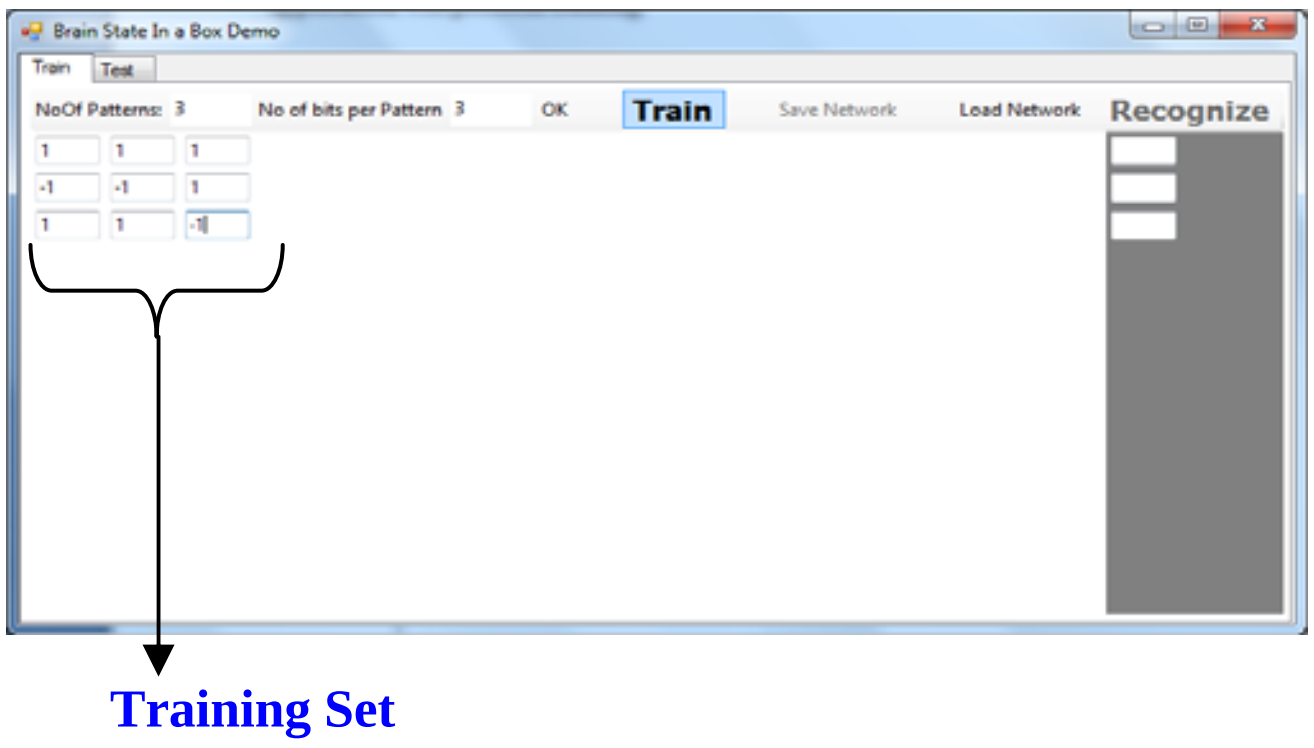

Fig. 9 Screen showing Training of BSB

Once the training has been completed it will display a message shown in figure 10.

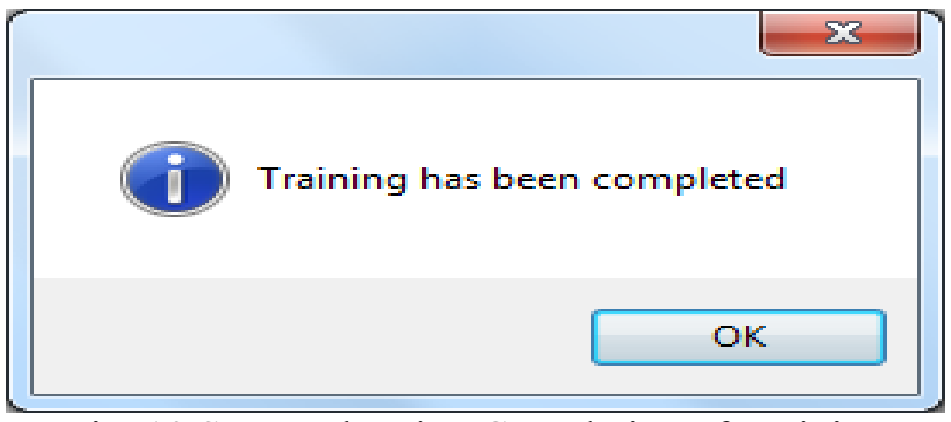

Fig. 10 Screen showing Completion of Training

## B. *Checking User authorization*

The application compares the output of given network with the password given by the user, if the stored user name and password matches with the given user name and password the user will be authenticated and the system resources will be allowed to the user.

If it does not match with the store password then the user will be restricted to access the server and its resources.

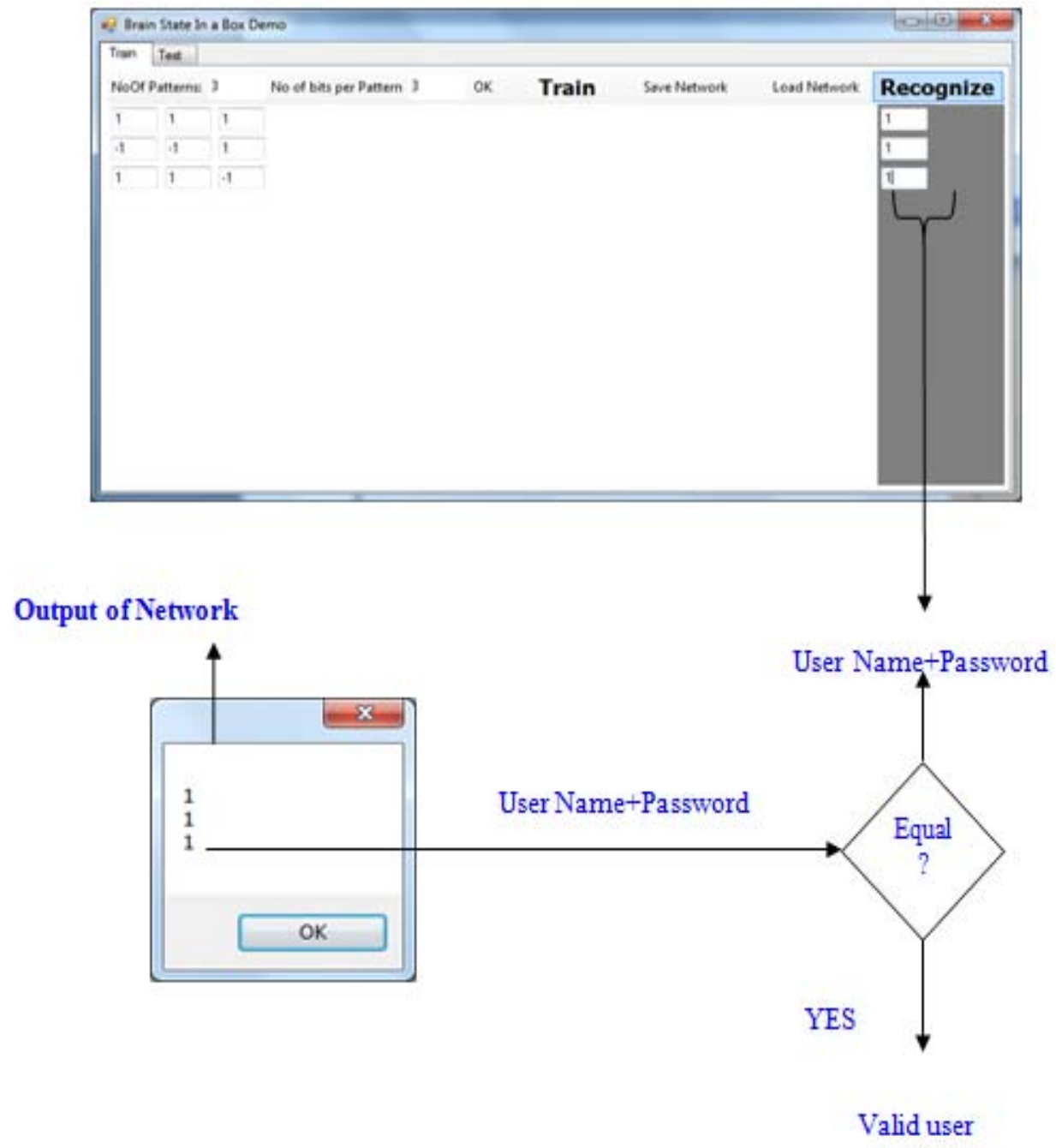

Fig. 11 Screen showing User Validation

## C. *BSB for Graphical Passwords*

This application cannot take an image directly as an input to the network. Before train the network, the required image should be converted in to text.

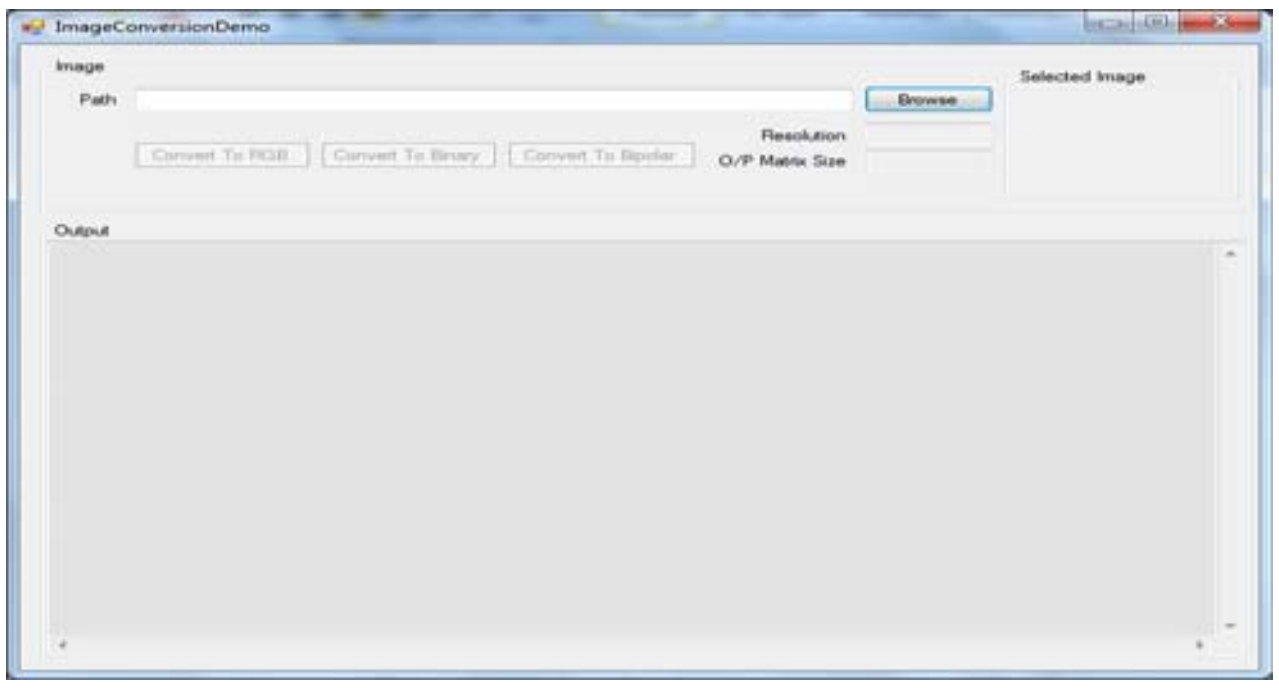

Fig. 12 Welcome screen for Image Authentication

In the screen shown in figure 12, *Path* indicates path of the image,*Selected Image* shows the selected image , *Resolution* specifies resolution of the selected image and O/P Matrix Size specifies size of the output matrix.

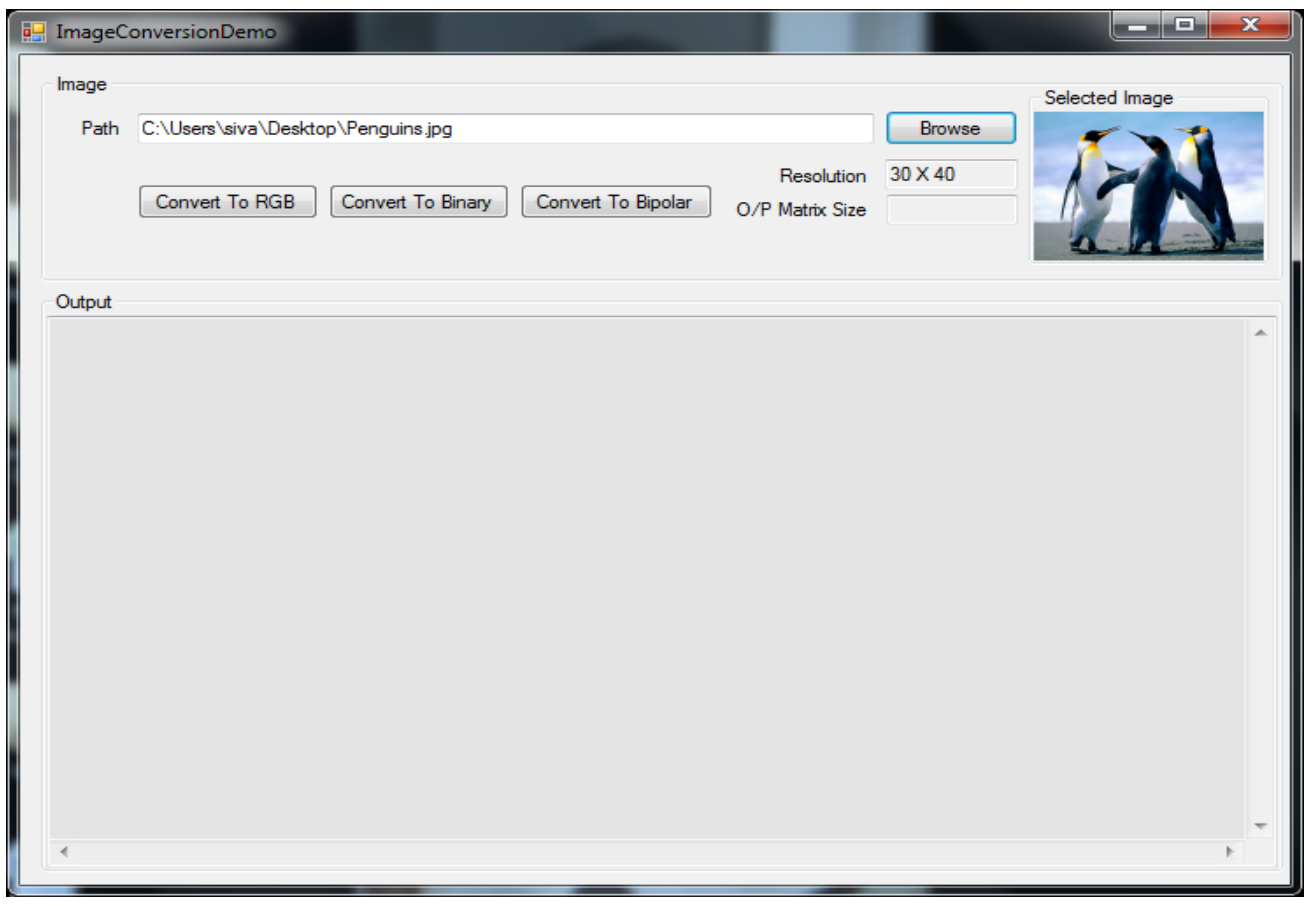

Fig. 13 Screen for loading the required image using BSB

Once the image has been selected *"Convert To RGB"button* is used to convert the image in to a matrix consisting of RGB values as shown in figure 14.

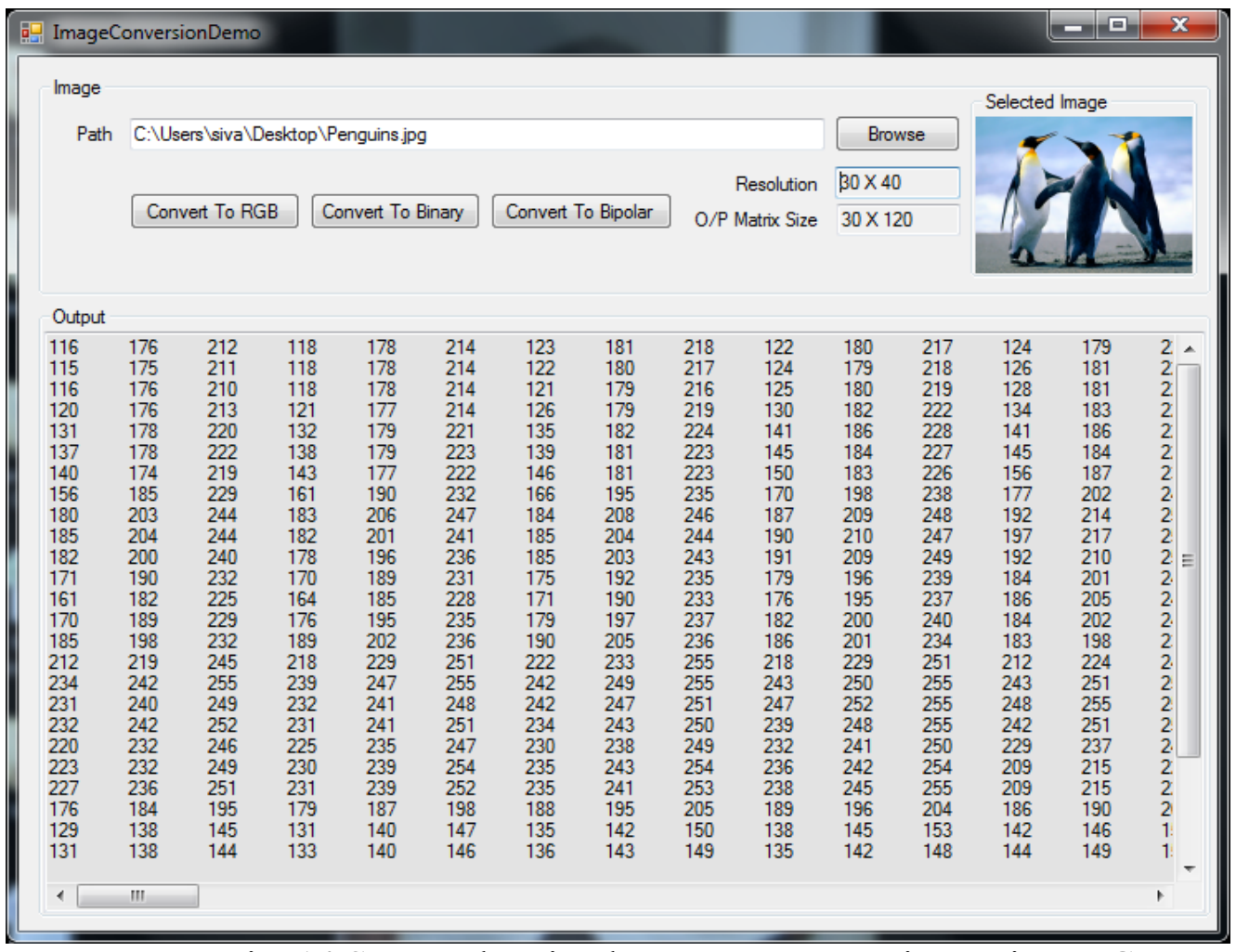

Fig. 14 Screen showing how to convert an image into RGB matrix

After selecting an image if "Convert To Binary" pressed in the application then output matrix is displayed in the output box as shown in the figure 15 .

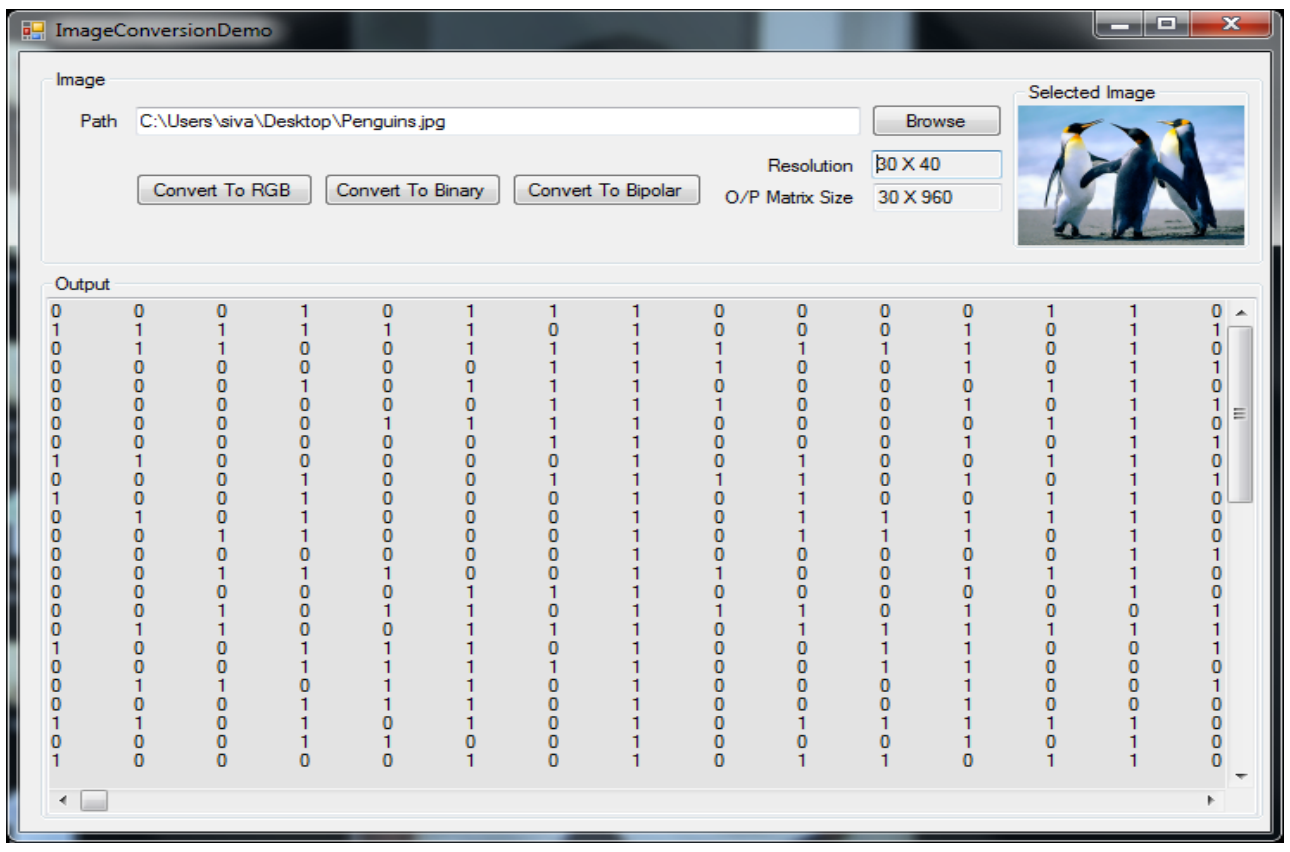

Fig. 15 Screen showing how to convert an image into a binary matrix

All the RGB values are converted in to binary values and they will be displayed in a matrix format. These values can be used as input for password authentication using BSB method and it satisfies the condition of taking probabilistic values as input for this method.

After specifying the path the required image will be retrieved and given as input for this method. When *"Convert To Bipolar"* is pressed the image will be converted in to bipolar values and it will be displayed as a matrix as in figure 16

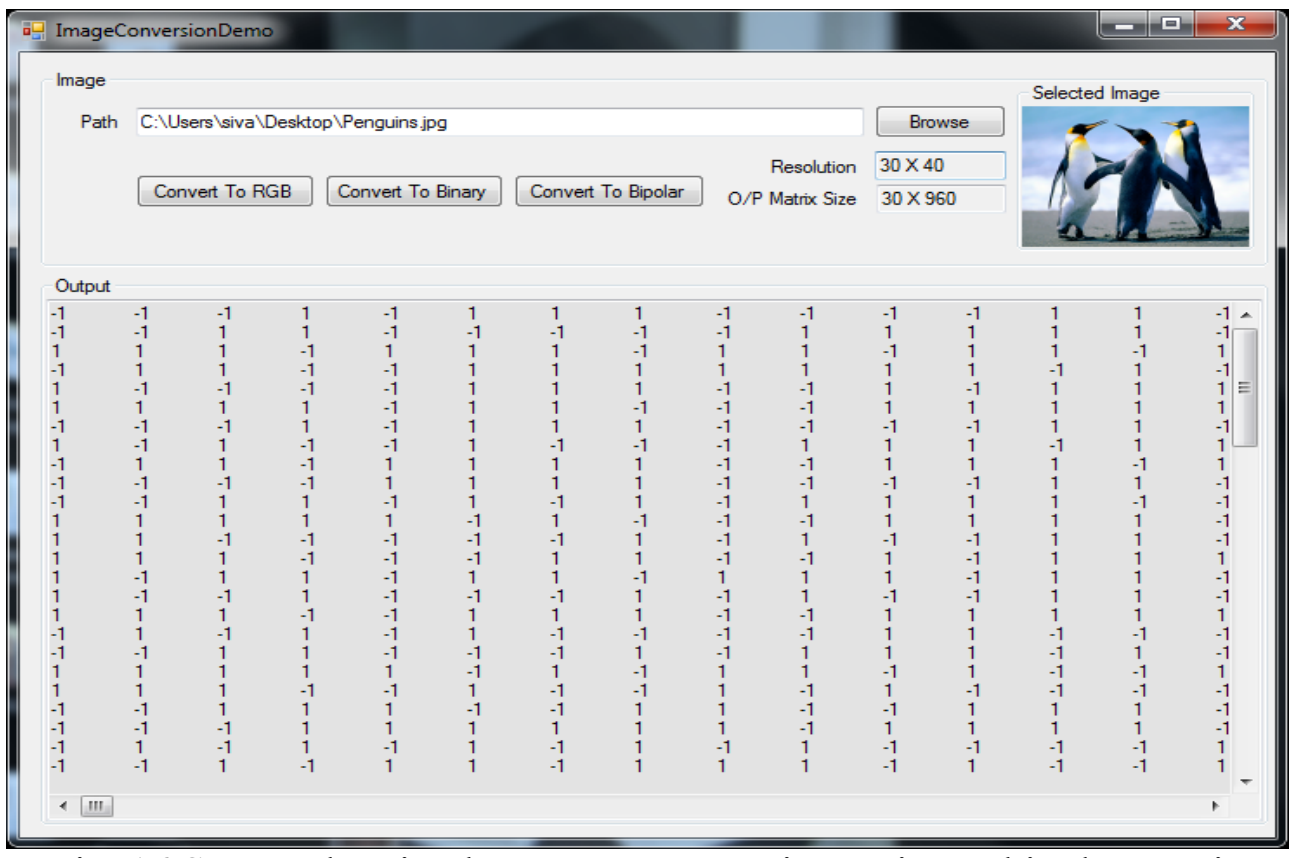

Fig. 16 Screen showing how to convert an image into a bipolar matrix

Once the image has been converted in to text, it can be used as normal textual password for giving it as input the BSB network.

## IV. CONCLUSION:

This paper introduced a password authentication using BSB. In this paper an algorithm for constructing the interconnection matrix W and vector b is proposed and implemented. This paper also provides a heuristic explanation for yielding an interconnection matrix with desired properties.
The desired properties include the asymmetry of W. The algorithm ensures that the negatives of the desired patterns are not automatically stored as asymptotically stable equilibrium points of the network, and it has provisions to minimize the number of spurious states. Digital computer simulations verified that our design algorithm yielded a network which stored all of the desired patterns as asymptotically stable equilibrium points with very few spurious states. The network has one main shortcoming: the network is not guaranteed to be globally stable.

## ACKNOWLEDGMENT

An assemblage of this nature could never have been attempted without reference to and inspiration from the works of others whose details are mentioned in reference section.

## AUTHORS

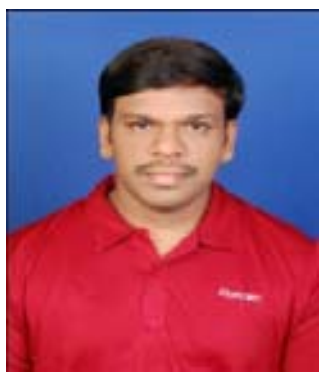

[1]A.S.N Chakravarthy received his M.Tech (CSE) from JNTU, Anantapur , Andhra Pradesh, India. Presently he is working as an Associate Professor in Dept. Of Computer Science and Engineering in Sri Aditya Engineering College, SuramPalem, AP, India. He is a research scholar under the supervision of Prof.P.S.Avadhani His research areas include Network Security, Cryptography, Intrusion Detection, Neural networks, Digital Forensics and Cyber Security.

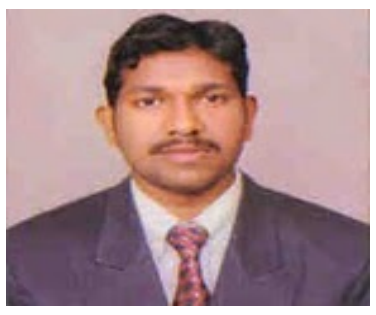

[2]Penmetsa V Krishna Raja received his M.Tech (CST) from A.U, Visakhapatnam, Andhra Pradesh, India. He is a research scholar under the supervision of Prof.P.S.Avadhani. His research areas include Network Security, Cryptography, Intrusion Detection, Neural networks.

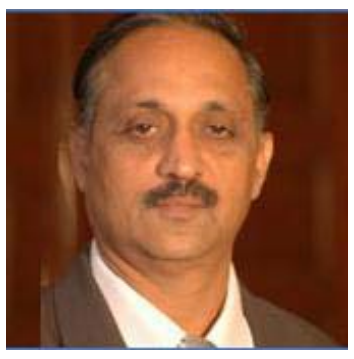

[3]Prof. P.S.Avadhani did his Masters Degree and PhD from IIT, Kanpur. He is presently working as Professor in Dept. of Computer Science and Systems Engineering in Andhra University college of Engg., in Visakhapatnam. He has more than 50 papers published in various National / International journals and conferences. His research areas include Cryptography, Data Security, Algorithms, and Computer Graphics, Digital Forensics and Cyber Security.